\documentclass[sn-nature]{sn-jnl}

\usepackage{graphicx}
\usepackage{multirow}
\usepackage{amsmath,amssymb,amsfonts}
\usepackage{amsthm}
\usepackage{mathrsfs}
\usepackage[title]{appendix}
\usepackage{xcolor}
\usepackage{textcomp}
\usepackage{manyfoot}
\usepackage{booktabs}
\usepackage{algorithm}
\usepackage{algorithmicx}
\usepackage{algpseudocode}
\usepackage{listings}
\usepackage{siunitx}
\usepackage[left]{lineno}

\begin{document}

\title{Magnetic Confinement of a Bubble of Supercooled $^3$He-A}

\author*[1]{\fnm{Luke} \sur{Whitehead}}\email{l.j.whitehead@lancaster.ac.uk}
\author[2]{\fnm{Andrew} \sur{Casey}}
\author[1]{\fnm{Richard P.} \sur{Haley}}
\author[2]{\fnm{Petri J.} \sur{Heikkinen}}
\author[2]{\fnm{Lev V.} \sur{Levitin}}
\author[1]{\fnm{Adam J.} \sur{Mayer}}
\author[2]{\fnm{Xavier} \sur{Rojas}}
\author[1]{\fnm{Tineke} \sur{Salmon}}
\author[2]{\fnm{John} \sur{Saunders}}
\author[1]{\fnm{Alex} \sur{Thomson}}
\author[1]{\fnm{Dmitry E.} \sur{Zmeev}}
\author[1]{\fnm{Samuli} \sur{Autti}}

\affil[1]{\orgdiv{Department of Physics}, \orgname{Lancaster University}, \orgaddress{ \city{Lancaster}, \postcode{LA1 4YB}, \country{UK}}}

\affil[2]{\orgdiv{Department of Physics}, \orgname{Royal Holloway, University of London}, \orgaddress{ \city{Egham}, \postcode{TW20 0EX}, \country{UK}}}

\abstract{We have designed and constructed a magnet surrounding a cylindrical volume of superfluid helium-3 to isolate a region of metastable, supercooled A-phase, entirely surrounded by bulk A-phase - isolating the 'bubble' from rough surfaces that can trigger the transition to the stable B-phase. We outline the design of the experimental cell and magnet, and show that the performance of the magnet is consistent with simulations, including the capability to producing the high field gradient required for generating a bubble. Future plans include the investigation of possible intrinsic mechanisms underpinning the A-B transition, with potential implications for early-universe cosmological phase transitions.}

\maketitle

\section{Introduction}\label{sec:introduction}

The exact process of nucleation of the B-phase of superfluid helium-3 from the supercooled A-phase is not well understood. Predictions made using homogeneous nucleation theory estimate a lifetime of the metastable state greater than the age of the observable universe \cite{Schiffer_Osheroff_1995}, owing to the small free energy difference between the phases and large surface tension of the interface separating them \cite{Osheroff_Cross_1977}. In practice, this transition is observed to happen within hours \cite{Schiffer_Osheroff_1995}.

The strong disagreement of the theory and experimental observations prompts investigation into the nucleation mechanisms that are responsible. It is well-established that external radiation sources \cite{Schiffer_O’Keefe_Hildreth_Fukuyama_Osheroff_1992} and surface roughness \cite{O’Keefe_Barker_Osheroff_1996} can reliably trigger the transition, but there could be the potential for intrinsic mechanisms to contribute, with experiments taking place in atomically smooth-walled nanofluidic cells to investigate this \cite{Heikkinen_Eng_Levitin_Rojas_Singh_Autti_Haley_Hindmarsh_Zmeev_Parpia_etal._2024}. We have designed a magnet surrounding a volume of superfluid helium-3 to create an isolated region (a 'bubble') of supercooled A-phase, completely surrounded by the stable A-phase, eliminating interactions of the bubble with the container surfaces which are known to induce the phase transition. The size of the region, as well as the pressure and temperature of the volume, can be varied to investigate their effects on the nucleation rate.

As the temperature approaches zero, superfluid helium-3 is essentially pure, making it an ideal candidate for studying nucleation theories of first-order phase transitions. The investigation of intrinsic mechanisms driving the A-B transition could have implications for early-universe cosmological phase transitions  - important for future space-based interferometers that could detect the gravitational waves produced by such an event \cite{Hindmarsh_Sauls_Zhang_Autti_Haley_Heikkinen_Huber_Levitin_Lopez-Eiguren_Mayer_etal._2024}. A new rapid intrinsic mechanism for example, would act to reduce the expected gravitational wave intensity \cite{Hindmarsh_Sauls_Zhang_Autti_Haley_Heikkinen_Huber_Levitin_Lopez-Eiguren_Mayer_etal._2024,Hindmarsh_Lüben_Lumma_Pauly_2021}.

\section{Experiment}\label{sec:experiment}

\begin{figure}[H]
    \centering
    \includegraphics[width=0.9\linewidth]{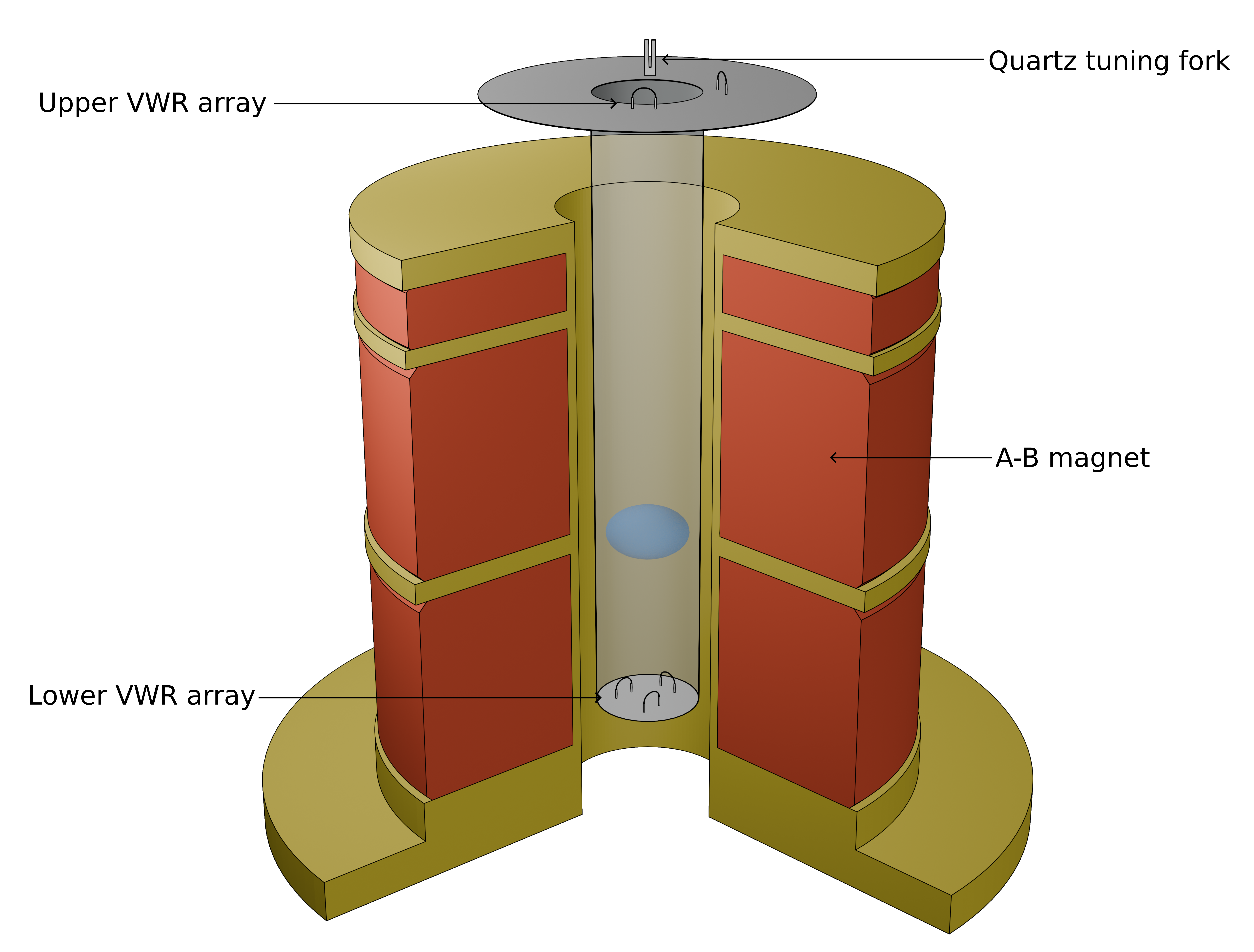}
    \caption{Schematic of the experimental cell and magnet. Above the main body of the inner cell but within the same volume are an array of VWRs and a quartz tuning fork for thermometry. The opposite end of the cell contains another series of VWRs for the same purpose. The A-B magnet used for generating the field gradient required to stabilise the bubble surrounds the helium-3 volume, and is shown in red and gold. The approximate location and size of the metastable A-phase bubble formed during the experiment is highlighted in blue, and is surrounded by regular A-phase. Not shown is the set of sintered-silver copper plates above the upper VWR array and tuning fork providing cooling for the sample, or the outer cell surrounding the experimental volume.}
    \label{fig:cell_schematic}
\end{figure}

The experimental volume (the 'inner cell') is a cylinder of Stycast 1266 epoxy reinforced paper, measuring 90 mm in length with an internal diameter of 14.9 mm, filled with $^3$He. The volume is in contact with silver-sintered copper plates which are cooled by nuclear demagnetisation, allowing the helium-3 to reach temperatures below \SI{200}{\micro\kelvin}. Surrounding this is another cylinder of helium-3 (the 'outer cell') acting as a thermal shield. Within the inner cell is a quartz tuning fork resonator \cite{Blaauwgeers_Blazkova_Človečko_Eltsov_deGraaf_Hosio_Krusius_Schmoranzer_Schoepe_Skrbek_etal._2007} and an array of vibrating wire resonators (VWRs), used as heaters and thermometers \cite{Guénault_Keith_Kennedy_Mussett_Pickett_1986} for the experiment, detailed in Figure \ref{fig:cell_schematic}. The VWRs are small loops of superconducting wire, driven to oscillate at their resonant frequency by the Lorentz force when an AC current is supplied. The damping experienced by the wire as it moves through the superfluid can be determined from the Faraday voltage, measured by a lock-in amplifier in parallel with the drive current. As the upper VWR array (closest to the copper refrigerant) is in the B-phase, the force measured by the wire is dependent on the quasiparticle density within the superfluid, allowing accurate measurement of the temperature within the experimental volume \cite{Enrico_Fisher_Watts-Tobin_1995}. To heat the superfluid, a VWR is driven above the pair-breaking velocity, creating quasiparticles that travel ballistically through the bulk $^3$He \cite{Lambert_1992}.

The magnet used to stabilise a region of magnetic field within the superfluid consists of three coils; two main opposing solenoids to produce the required high field gradient and one smaller compensation coil. All are constructed from a single length of copper clad NbTi superconducting wire measuring \SI{279}{\micro\meter} in diameter, with each of the larger coils consisting of 8004 turns across 85 layers, and the smaller 2106 turns across 88 layers. The compensation coil ensures the field generated by the magnet does not interfere with the demagnetisation field required for cooling. The brass former for the magnet is attached to the inside of a radiation shield surrounding much of the dilution refrigeration unit and the demagnetisation stage of the cryostat, thermally anchoring the magnet to the still plate at approximately \SI{0.5}{\kelvin}.

\begin{figure}[H]
    \centering
    \includegraphics[width=0.95\linewidth]{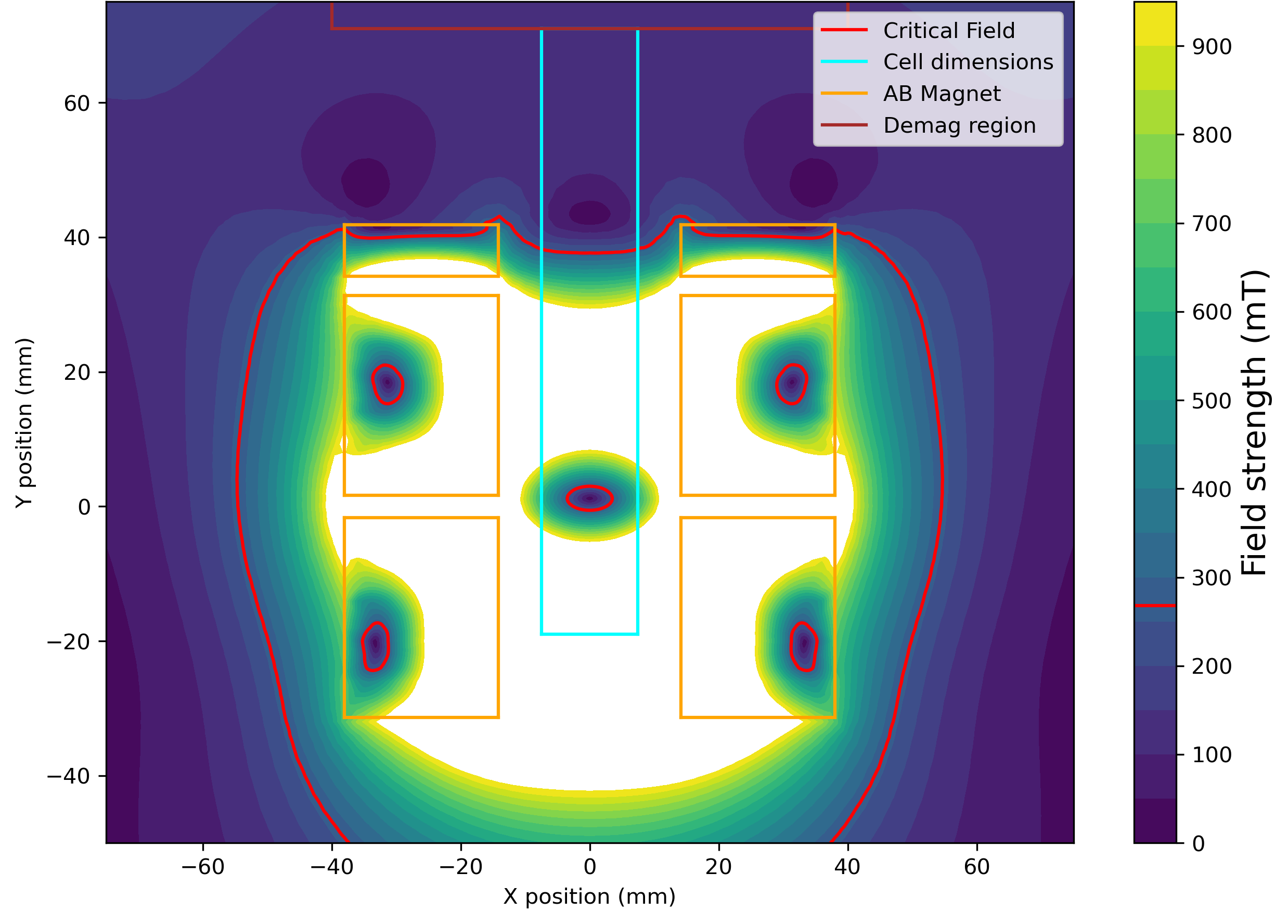}
    \caption{Simulated magnetic field profile. Outline of the experimental cell (blue) and magnet (orange) are shown. The critical field for creating an A-B boundary (red) is given for typical experimental conditions - in this case, a temperature of 0.75 $T/T_C$ and a pressure of \SI{20}{\bar}, resulting in a region in the cell where B-phase is preferred, surrounded by a volume of A-phase. Regions where the field exceeds \SI{900}{\milli\kelvin} are uncoloured.}
    \label{fig:field_plot}
\end{figure}

Figure \ref{fig:field_plot} shows the simulated magnetic field profile in a typical experiment scenario, with the critical field region highlighted.

\begin{figure}[H]
    \centering
    \includegraphics[width=0.95\linewidth]{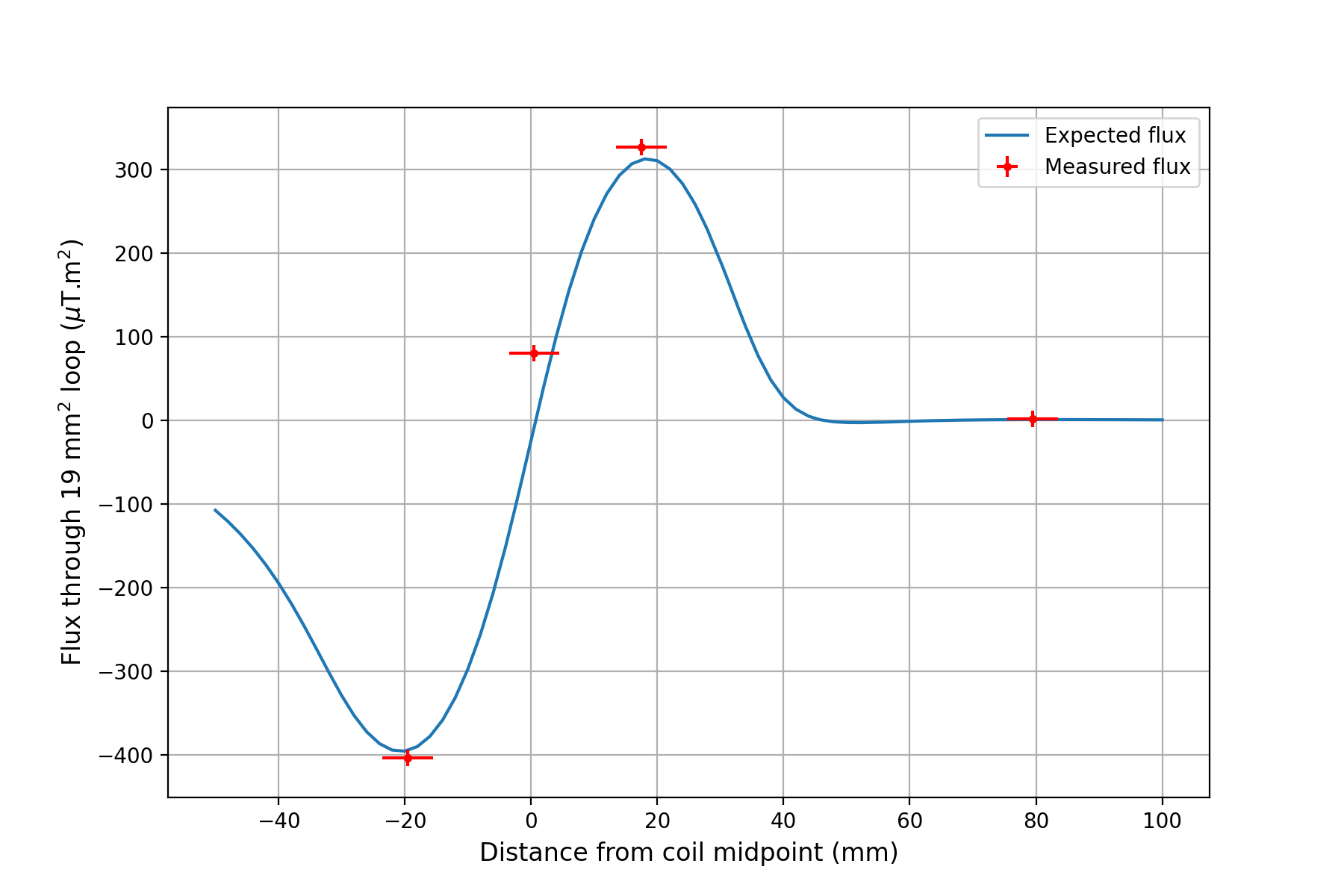}
    \caption{Predicted (blue) and measured (red) magnetic flux along the length of the magnet bore. Measured points from left to right correspond to the expected values for the field minimum, coil midpoint, field maximum and far above the coil.}
    \label{fig:flux_measurement}
\end{figure}

A probe made up of a series of detection coils was developed to measure the magnetic field at four critical points: the expected maxima, the coil midpoint, and far away from the main volume, within the demagnetisation region. At all points the measured field was within 5\% of predictions (see Figure \ref{fig:flux_measurement}), including complete compensation of the primary A-B solenoids in the demagnetisation region.

If we assume the radial component of the magnetic field produced by the magnet also has a $\pm 5 \%$ uncertainty, then the radius of the bubble will likewise be within 5\% of the predicted value. A typical bubble with a radius of 5-6 mm can therefore be created to within 0.3 mm.

The magnet has been confirmed to operate above the required current threshold of \SI{15}{\ampere} for stabilising the A-phase, reaching \SI{19}{\ampere} before quenching, and can run simultaneously with the main demagnetisation magnet without affecting its field.

\section{Discussion}\label{sec:discussion}

We have developed a magnet capable of stabilising an isolated region of metastable superfluid helium-3 away from container walls, with initial testing showing the predicted magnetic field can be stabilised under planned experimental conditions. Previous measurements have shown how the presence of an A-B boundary provides a thermal resistance for quasiparticles \cite{Bradley_Fisher_Guénault_Haley_Martin_Pickett_Roberts_Tsepelin_2006a,Bradley_Fisher_Guénault_Haley_Martin_Pickett_Roberts_Tsepelin_2006b}. By generating quasiparticles at one end of the cell using one resonator array and measuring the corresponding temperature rise at the other, we can determine the phase contained in the bubble, and therefore the lifetime of the metastable A-phase. In addition, the latent heat released when the bubble undergoes the transition from the supercooled A-phase to the B-phase can be used \cite{Bradley_Fisher_Guénault_Haley_Martin_Pickett_Roberts_Tsepelin_2006a}.

Upcoming experiments will allow us to investigate potential intrinsic nucleation mechanisms that trigger the transition from the supercooled A-phase to the B-phase by systematically varying the bubble size, temperature and pressure and observing the time taken for the bubble to transition. 

\section*{Acknowledgements}
This work was funded by UKRI EPSRC and STFC (Grants ST/T006773/1, EP/W015730/1), as well the European Union’s Horizon 2020 Research and Innovation Programme under Grant Agreement no 824109 (European Microkelvin Platform).

\bibliography{sn-bibliography}

\end{document}